\documentclass[iop]{emulateapj}

\def\lapp{\ifmmode\stackrel{<}{_{\sim}}\else$\stackrel{<}{_{\sim}}$\fi}
\def\gapp{\ifmmode\stackrel{>}{_{\sim}}\else$\stackrel{>}{_{\sim}}$\fi}
\usepackage{multirow}
\usepackage{color}
\usepackage{amsmath}
\usepackage{soul}
\usepackage{lineno}

\begin{document}

\title{Contemporaneous broadband observations of three high-redshift BL Lac Objects}

\author{
M.~Ackermann\altaffilmark{1},
M.~Ajello\altaffilmark{2},
H.~An\altaffilmark{3,4},
L.~Baldini\altaffilmark{5,3},
G.~Barbiellini\altaffilmark{6,7},
D.~Bastieri\altaffilmark{8,9},
R.~Bellazzini\altaffilmark{10},
E.~Bissaldi\altaffilmark{11},
R.~D.~Blandford\altaffilmark{3},
R.~Bonino\altaffilmark{12,13}
% ...
}
\altaffiltext{1}{Deutsches Elektronen Synchrotron DESY, D-12738 Zeuthen, Germany}
\altaffiltext{2}{Department of Physics and Astronomy, Clemson University, Kinard Lab of Physics, Clemson, SC 29634-0978, USA}
\altaffiltext{3}{W. W. Hansen Experimental Physics Laboratory, Kavli Institute for Particle Astrophysics and Cosmology, Department of Physics and SLAC National Accelerator Laboratory, Stanford University, Stanford, CA 94305, USA}
\altaffiltext{4}{email: hjans@stanford.edu}
\altaffiltext{5}{Universit\`a di Pisa and Istituto Nazionale di Fisica Nucleare, Sezione di Pisa I-56127 Pisa, Italy}
\altaffiltext{6}{Istituto Nazionale di Fisica Nucleare, Sezione di Trieste, I-34127 Trieste, Italy}
\altaffiltext{7}{Dipartimento di Fisica, Universit\`a di Trieste, I-34127 Trieste, Italy}
\altaffiltext{8}{Istituto Nazionale di Fisica Nucleare, Sezione di Padova, I-35131 Padova, Italy}
\altaffiltext{9}{Dipartimento di Fisica e Astronomia ``G. Galilei'', Universit\`a di Padova, I-35131 Padova, Italy}
\altaffiltext{10}{Istituto Nazionale di Fisica Nucleare, Sezione di Pisa, I-56127 Pisa, Italy}
\altaffiltext{11}{Istituto Nazionale di Fisica Nucleare, Sezione di Bari, I-70126 Bari, Italy}
\altaffiltext{12}{Istituto Nazionale di Fisica Nucleare, Sezione di Torino, I-10125 Torino, Italy}
\altaffiltext{13}{Dipartimento di Fisica Generale ``Amadeo Avogadro" , Universit\`a degli Studi di Torino, I-10125 Torino, Italy}
% ...

\begin{abstract}
        We have collected broadband spectral energy distributions (SEDs) of three
BL Lac objects, 3FGL~J0022.1$-$1855 ($z$=0.689),
3FGL~J0630.9$-$2406 ($z\gapp$1.239), and 3FGL~J0811.2$-$7529 ($z$=0.774),
detected by {\it Fermi} with relatively flat GeV spectra.
By observing
simultaneously in the near-IR to hard X-ray band, we can well characterize the high
end of the synchrotron component of the SED. Thus, fitting the SEDs to synchro-Compton
models of the dominant emission from the relativistic jet, we can constrain the
underlying particle properties and predict the shape of the GeV Compton component.
Standard extragalactic background light (EBL) models explain
the high-energy absorption well, with poorer fits
for high UV models. The fits show clear evidence for
EBL absorption in the {\it Fermi} spectrum
of our highest redshift source 3FGL~J0630.9$-$2406.
While synchrotron self-Compton models adequately describe the SEDs,
the situation may be complicated by possible external Compton components.
For 3FGL~J0811.2$-$7529, we also discover a nearby serendipitous source in the X-ray data,
which is almost certainly another lower synchrotron peak frequency ($\nu_{\rm pk}^{\rm sy}$) BL Lac,
that may contribute flux in the {\it Fermi} band.
Since our sources are unusual high-luminosity,
moderate $\nu_{\rm pk}^{\rm sy}$ BL Lacs we compare these quantities and
the Compton dominance, the ratio of peak inverse-Compton to peak synchrotron luminosities
($L^{\rm IC}_{\rm pk}/L^{\rm sy}_{\rm pk}$),
with those of the full {\it Fermi} BL Lac population.
\end{abstract}


\section{Introduction}
Blazars, active galactic nuclei (AGN) with strong nonthermal
emission from an aligned
relativistic jet \citep[][]{br78,up95},
are the most luminous persistent objects in the universe.
These sources emit photons
across the whole electromagnetic spectrum from the radio to gamma-ray bands.
Their spectral energy distributions (SEDs) are well characterized with a
double-hump structure where the low-energy hump, peaking in the IR/optical/UV/X-ray
band, is thought to be produced by synchrotron emission of the jet electrons.
Their high-energy peak in the gamma-ray band is produced by
synchrotron self-Compton (SSC) and external Compton (EC) scattering, or possibly
by hadronic processes \citep[e.g.,][]{mb92,bms97,gtf+10}.

Blazars are heuristically classified into flat spectrum radio quasars (FSRQs) and
BL Lacertae objects (BL Lacs). The former show broad optical emission lines associated
with clouds surrounding or in the accretion disk. The latter lack such lines and have a jet
continuum strong enough to obscure spectral features of the host galaxy
\citep[][]{mbi+96,lpp+04}.  \citet{pg95} further divided BL Lacs
based on the synchrotron peak frequency ($\nu^{\rm sy}_{\rm pk}$) into low synchrotron peak
(LSP, $\nu^{\rm sy}_{\rm pk}<10^{14}\rm Hz$), intermediate peak
(ISP, $10^{14}\rm Hz<\nu^{\rm sy}_{\rm pk}<10^{15}\rm Hz$), and high peak
(HSP, $10^{15}\rm Hz<\nu^{\rm sy}_{\rm pk}$) subclasses. FSRQs are
almost all classified as LSP \citep[][]{fermiblazar10}.

        \citet{fmc+98} found that 5\,GHz luminosity,
the synchrotron peak luminosity ($L^{\rm sy}_{\rm pk}$), and
the gamma-ray dominance (ratio of the peak
gamma-ray to peak synchrotron $\nu F_{\nu}$ luminosity)
are correlated with $\nu^{\rm sy}_{\rm pk}$.
They characterize this as
a ``blazar sequence'' trend from low-peaked powerful sources (i.e., FSRQs) to high-peaked
less powerful sources (HSPs). A plausible physical explanation for this
sequence is provided by \citet{gcf+98}; more luminous sources tend to have stronger disk
accretion, and the external photons from the broad line region (BLR) or the disk in these
sources provide additional seeds for Compton upscattering which cools the jet
electrons, lowering $\nu^{\rm sy}_{\rm pk}$, while increasing the Compton luminosity.
Indeed, as the typical accretion state evolves over cosmic time, this picture
may provide an explanation of evolution in the FSRQ/BL Lac blazar populations
\citep[][]{bd02,cd02}. Quantitatively, this may explain the apparent ``negative evolution''
(increase at low redshift) observed for HSP BL Lacs \citep[][]{rsp+00,beb+03,arg+14}.

\newcommand{\marka}{\tablenotemark{a}}
\newcommand{\markb}{\tablenotemark{b}}
\newcommand{\markc}{\tablenotemark{c}}
\begin{table*}[t]
\vspace{-0.0in}
\begin{center}
\caption{Summary of observations used in this work
\label{ta:ta1}}
\vspace{-0.05in}
\scriptsize{
\begin{tabular}{cccccccc} \hline\hline
Source & R.A. &  Decl. & Redshift & Observatory        & Start date   & Obs. ID  & Exposure      \\
  &    &  &   &   & (MJD)       &       & (ks)              \\ \hline
\multirow{4}{*}{J0022} & \multirow{4}{*}{0$^{\rm h}$22$^{\rm m}$09.25$^{\rm s}$} & \multirow{4}{*}{$-$18$^\circ$53$'$34.9$''$} & \multirow{4}{*}{0.774}  & GROND   & 57031.1 & $\cdots$ & 0.25/0.24\marka \\
                       & & & & {\em Swift}   & 57031.7 & 00080777001 & 1.9\markb \\
                       & & & & {\em XMM}     & 57026.8 & 0740820501 & 15/9\markc \\
                       & & & & {\em NuSTAR}  & 57026.7 & 60001141002--4 & 110    \\ \hline
\multirow{4}{*}{J0630} & \multirow{4}{*}{6$^{\rm h}$30$^{\rm m}$59.515$^{\rm s}$} &\multirow{4}{*}{$-$24$^\circ$06$'$46.09$''$} &\multirow{4}{*}{$>$1.239} & GROND   & 56949.2 & $\cdots$ & 0.25/0.24\marka \\
                       & & & & {\em Swift}   & 56948.5 & 00080776001 & 0.27\markb \\
                       & & & & {\em XMM}     & 56948.2 & 0740820401 & 8/4\markc \\
                       & & & & {\em NuSTAR}  & 56947.7 & 60001140002 & 67  \\ \hline
\multirow{4}{*}{J0811} & \multirow{4}{*}{8$^{\rm h}$11$^{\rm m}$03.214$^{\rm s}$} & \multirow{4}{*}{$-$75$^\circ$30$'$27.85$''$} & \multirow{4}{*}{0.689} & GROND   & 56903.3 & $\cdots$ & 0.25/0.24\marka \\
                       & & & & {\em SWIFT}   & 56908.2 & 00091903001 & 0.39\markb \\
                       & & & & {\em XMM}     & 56901.2 & 0740820601 & 9/6\markc \\
                       & & & & {\em NuSTAR}  & 56901.2 & 60001142002 & 113  \\ \hline
\end{tabular}}
\end{center}
\hspace{-2.0 mm}
$^{\rm a}${ For {\it g$'$r$'$i$'$z$'$/JHK} bands.}\\
$^{\rm b}${ For the UW1 band. Exposures in the other UVOT bands may differ from this value.}\\
$^{\rm c}${ For MOS1,2/PN.}\\
\vspace{-1.0 mm}
\end{table*}

           On the other hand, \citet{gpp+12} used Monte Carlo simulations to argue that the
$L^{\rm sy}_{\rm pk}$ and $\nu^{\rm sy}_{\rm pk}$ anti-correlation may be primarily a
selection effect.  \citet{pgr12} discuss four sources with high $\nu^{\rm sy}_{\rm pk}$ and
high peak (synchrotron + SSC) power as examples well off of the blazar sequence.
Such sources might be FSRQs with unusually strong jet emission along the Earth line-of-sight
masking the underlying host components. Thus simultaneous observations and careful SED
modeling of such (generally higher-redshift) BL Lac sources is interesting as it can
help us understand the underlying emission zone physics and whether it is truly different
from the bulk of the blazar population. Characterization via less redshift-dependent parameters
\citep[e.g. gamma-ray dominance or Compton dominance;
see][for example]{fmc+98,f13} may also help clarify
their place in the population. Also, comparing robust SED model fits with
gamma-ray spectra of high-$z$ blazars can reveal the effect of absorption
by the extragalactic background light (EBL), which provides important constraints
on evolution of cosmic star formation \citep[e.g.,][]{fermiEBL, HESSEBL}.
BL Lacs are believed to have higher Compton dominance and less sensitivity
to local soft photon fields and so are particularly useful for such study.

        Appropriate high-redshift HSP BL Lac objects are rare because they
are faint especially in the gamma-ray band, and HSPs appear to exhibit negative
evolution \citep[][]{arg+14}. We select three
{\it Fermi}-detected \citep[][]{fermi2fgl,fermi2lac} sources,
3FGL~J0022.1$-$1855 (J0022, $z=0.774$),
3FGL~J0630.9$-$2406 (J0630, $z>1.239$),
and 3FGL~J0811.2$-$7529 (J0811, $z=0.689$), whose optical spectra
are unusual, showing no emission lines but a set of strong low excitation (Mg~I,
Fe~II, Al~II etc) absorption lines on a blue, power-law
continuum. These indicate that the AGN is viewed through the disk of an intervening
absorber. In \citet{src+13}, this was taken to be the host galaxy; indeed for
J0630 the photometry of \citet{rsg+12} supports this as the host redshift. With
estimated redshifts of  0.774, $>$1.239, and 0.689 \citep[][]{rsg+12,src+13} for
J0022, J0630 and J0811, respectively,
these are thus luminous high-peak sources suitable for studying the extreme of the BL Lac
population. At these redshifts, we may also see the effects of
extragalactic background light absorption at
the high end of the {\it Fermi} band. To probe this absorption, and the high end
of the jet particle population most sensitive to Compton cooling, we require particularly
good characterization of the peak and high-energy cutoff (near-IR to hard X-ray)
of the synchrotron component. Under classic SSC modeling,
this allows us to characterize the high-energy Compton component, as well, thus
providing inferences about the Compton cooling at the source and EBL absorption of the
GeV photons as they propagate to Earth.

In this paper, we present broadband SEDs of the three high-redshift BL Lacs which are
simultaneous across the critical $\nu >\nu^{\rm sy}_{\rm pk}$ range (Section~\ref{sec:sec2}).
J0630 has been previously discussed as a high-$\nu^{\rm sy}_{\rm pk}$,
high-power source \citep{pgr12}; our improved data allow more refined modeling,
which is discussed in Section~\ref{sec:sec3}, including EBL constraints. The implications
of our inferred model parameters are discussed in Section~\ref{sec:sec4}.
We use $H_0=70\rm \ km\ s^{-1}\ Mpc^{-1}$, $\Omega_m=0.3$, $\Omega_{\Lambda}=0.7$ \citep[e.g.,][]{ksd+11},
and redshift values given in Table~\ref{ta:ta1} ($z=1.239$ for J0630) throughout.

\section{Observations and Data Reduction}
\label{sec:sec2}

BL Lac objects can be variable on all timescales from minutes to years \citep[][]{HESSvariable},
so coordinated broad-band coverage is important for characterizing the instantaneous SED.
We therefore carried out nearly contemporaneous observations of the sources using the
Gamma-Ray burst Optical/Near-Infrared Detector (GROND)
instrument at the 2.2-m MPG telescope at the ESO La Silla Observatory \citep{gbc+08}
as well as the {\it Swift} \citep{gcg+04}, {\it XMM-Newton} \citep{jla+01}
and {\it NuSTAR} \citep{hcc+13} satellites, covering the upper range of the synchrotron component.
Our sources showed relatively modest variability in the {\it Fermi} \citep[][]{fermimission} band
and so we average over 6 years of Large Area Telescope (LAT) data to best characterize the
mean Compton component of these relatively faint (but luminous, for BL Lacs) sources.
Archival radio, optical, and near-IR observations are provided for comparison
although we do not use them in the SED fitting.

\subsection{Contemporaneous observations: GROND, Swift, XMM-Newton, and NuSTAR}
\label{sec:sec2_1}

The GROND data were reduced and analyzed with the standard tools and methods
described in \citet{kkg+08}. The photometric data were obtained using
FWHM-matched PSF ($g^\prime r^\prime i^\prime z^\prime$) or aperture photometry ({\it JHK}).
The $g^\prime$, $r^\prime$, $i^\prime$, and $z^\prime$ photometric
calibration was obtained via standard star fields observed on the
same nights as the target integrations. The {\it J}, {\it H}, and {\it Ks} photometry was
calibrated against selected in-field 2MASS stars \citep[][]{scs+06}.

For {\it Swift} UVOT data, we performed aperture photometry for the six
{\it Swift} filters \citep[][]{pbp+08} using the {\tt uvotsource} tool
in HEASOFT 6.16\footnote{http://heasarc.nasa.gov/lheasoft/}. We measured photometric magnitude of the sources using
a $R=5^{\prime\prime}$ aperture. Backgrounds were estimated using a $R=20^{\prime\prime}$ circle
near the source.

X-ray SEDs of the sources were measured with {\it XMM-Newton} and {\it NuSTAR}.
The sources were detected with very high significance ($>20\sigma$)
with {\it XMM-Newton} but with relatively low significance ($\gapp 6\sigma$)
with {\it NuSTAR}.
For the {\em XMM-Newton} data, we processed the observation data files
with {\ttfamily epproc} and {\ttfamily emproc} of Science Analysis System (SAS)
version 14.0.0\footnote{http://xmm.esac.esa.int/sas/} and then applied
standard filters. The {\it NuSTAR} data were processed with the standard pipeline
tool {\tt nupipeline} of {\tt nustardas}
1.4.1 integrated in the HEASOFT 6.16.
We used {\it NuSTAR} CALDB version 20140414 and applied the standard
filters.\footnote{See http://heasarc.gsfc.nasa.gov/docs/nustar/analysis/nustar\\\_swguide.pdf for more details}
We then extracted source events using circular regions with $R=20^{\prime\prime}$ and $R=30^{\prime\prime}$
for the {\it XMM-Newton} and the {\it NuSTAR} data, respectively. Backgrounds were
extracted from nearby source-free regions.

\subsection{Gamma-ray observations}
\label{sec:sec2_2}
\begin{figure}
\includegraphics[width=3.65 in]{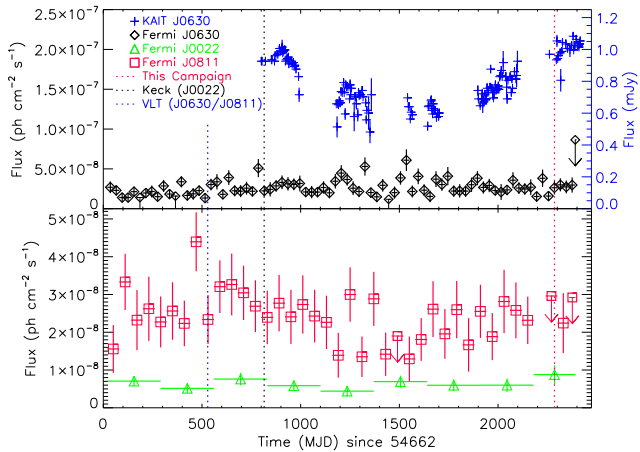}
\vspace{-3mm}
\figcaption{Optical ({\it R} band) and gamma-ray (100\,MeV--500\,GeV) light curves.
The top panel shows KAIT (right scale)
and {\it Fermi} (left scale) fluxes for J0630. Our contemporaneous observation
epoch and the optical spectrum epochs are marked. The lower panel shows the LAT light curves and
multiwavelength epochs for J0022 and J0811. The modest LAT variability justifies
the use of mission-averaged spectra.
\label{fig:fig1}
}
\vspace{0mm}
\end{figure}

For the gamma-ray data, we used the {\it Fermi} observations taken between
2008 August 4 and 2015 January 31. The Pass 8 data \citep[][]{fermiP8}, based on a complete
and improved revision of entire LAT event-level analysis, were downloaded
from Fermi Science Support Center\footnote{http://fermi.gsfc.nasa.gov/ssc/},
and we analyzed the data using the {\it Fermi}
Science tool 10-00-04 along with the instrument response functions (IRFs) P8R2\_SOURCE\_V6.
We extracted source class events in the 100\,MeV--500\,GeV band in a
$R=5^\circ$ region of interest (ROIs)
and $<80^\circ$ zenith angle and $<52^\circ$ rocking angle cuts.
These events were analyzed using the background models ({\tt gll\_iem\_v06} and
{\tt iso\_P8R2\_SOURCE\_V6\_v06}) and all 3FGL sources within $15^\circ$.
We first modeled fluxes on a one-month cadence to check for strong source
variability using the standard {\it Fermi} likelihood analysis with {\tt gtlike}
(see Figure~\ref{fig:fig1} and Section~\ref{sec:sec3_1}).
No strong flares were seen and so we combined all the LAT
data, modeling the mission-averaged spectrum.
In Figure~\ref{fig:fig1}, we mark the epochs of the
contemporaneous campaign and historical spectra. For J0630 we also have access
to optical monitoring from the KAIT program \citep[][]{crf+14}, shown on the top panel.
Variability is clearly seen in the optical band.

\subsection{Archival observations}
\label{sec:sec2_3}
For comparison, we also collected archival data in the radio to UV band. We assembled data
from various catalogs (e.g., {\it WISE} and 2MASS for IR data) or reanalyze the
archival data (e.g., VLT/Keck spectra and {\it Swift} UVOT). For the catalog data, we convert the
magnitude to flux appropriately.
The VLT/Keck data reduction and calibration were described in \citet{src+13}.
The archival UVOT data are processed as described above (Section~\ref{sec:sec2_1}).
The measurements are corrected for Galactic extinction in constructing
the SED (Section~\ref{sec:sec3_2}).
Archival measurements are used only in flux variability studies.

\subsection{Discovery of a serendipitous source}
\label{sec:sec2_4}

\begin{figure*}
\centering
\begin{tabular}{ccc}
\hspace{-3.0 mm}
\includegraphics[width=2.1 in]{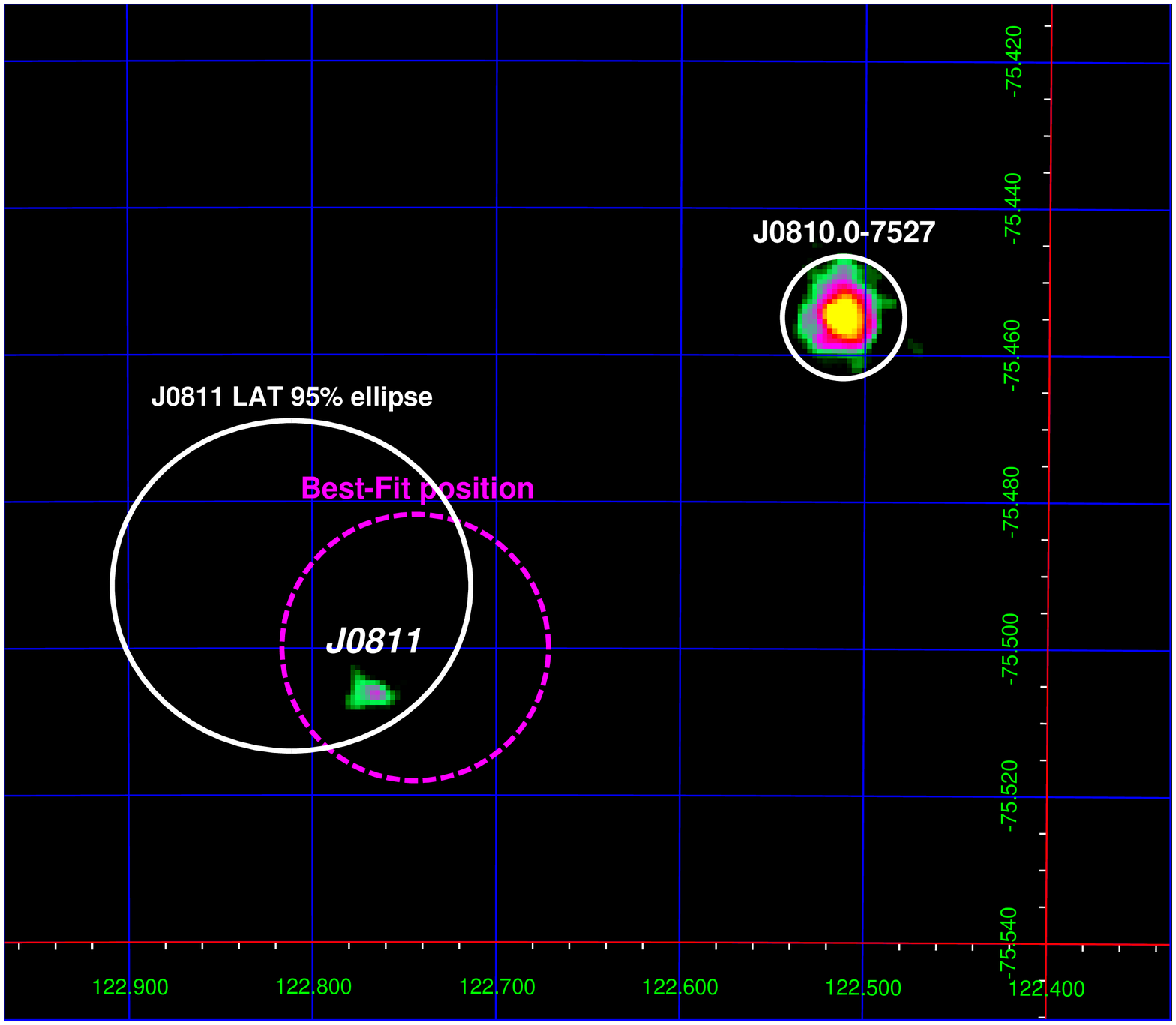} &
\hspace{1.0 mm}
\includegraphics[width=2.05 in]{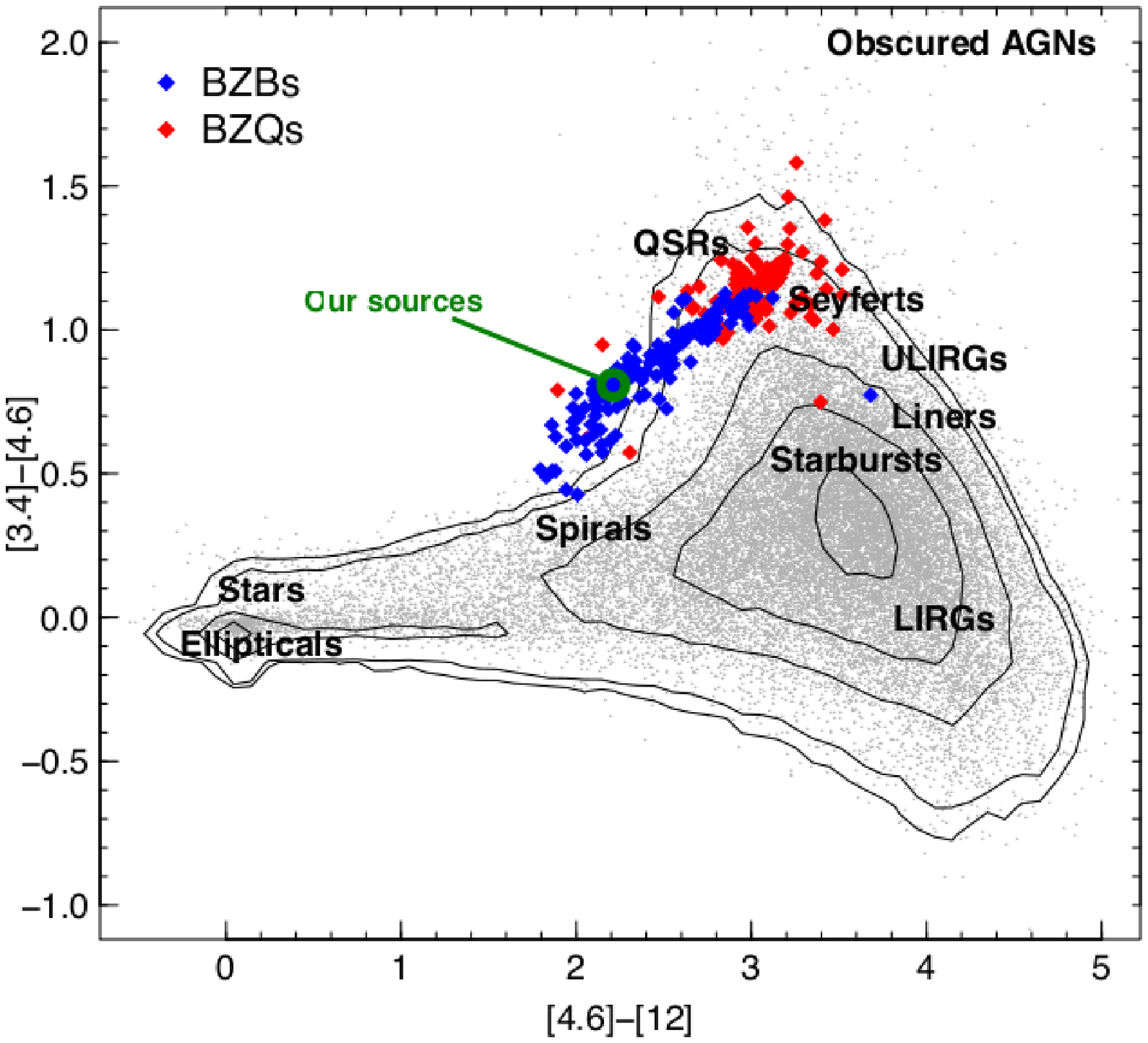} &
\hspace{-3.0 mm}
\includegraphics[width=2.7 in]{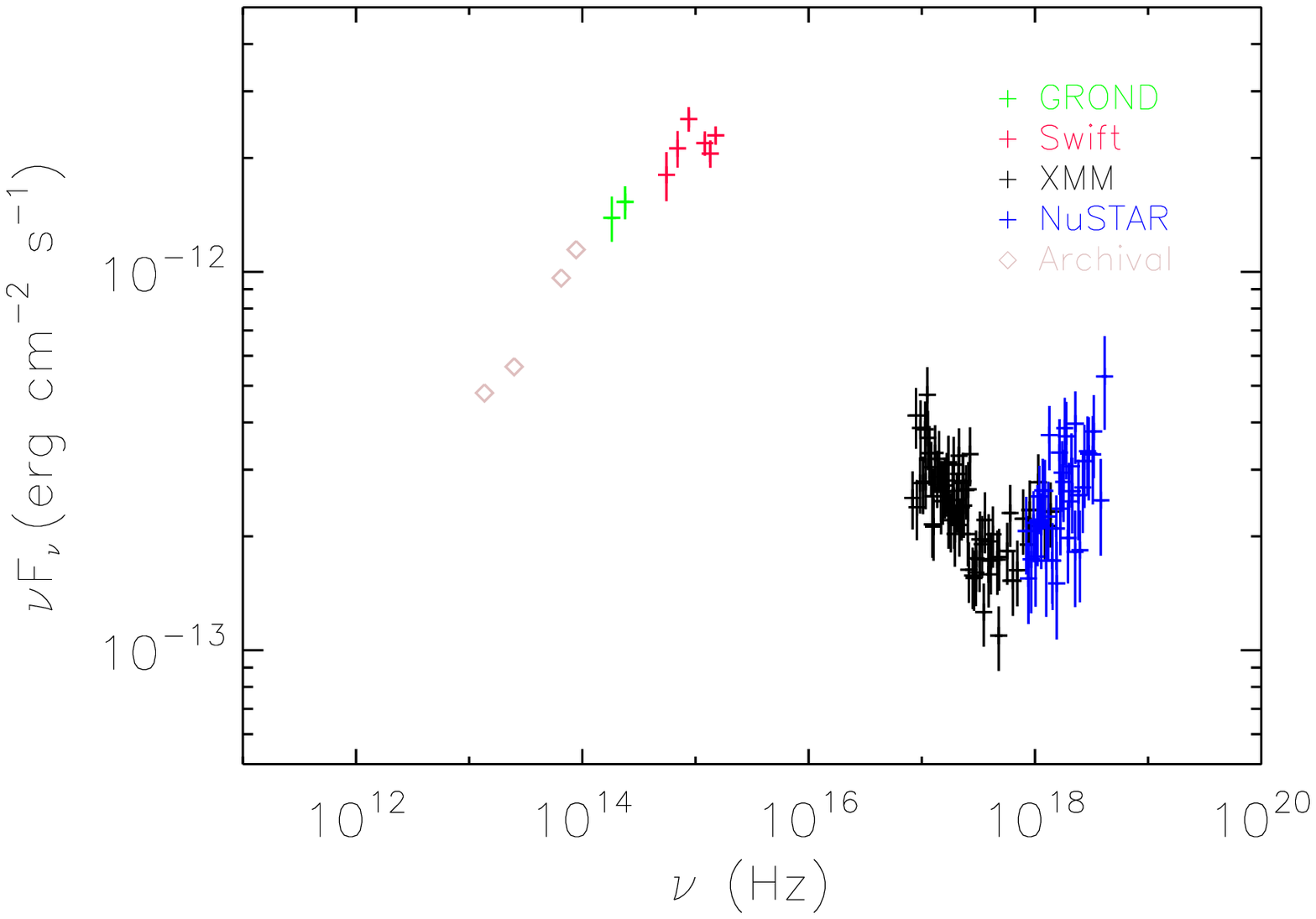} \\
\end{tabular}
\figcaption{{\it Left}: {\it NuSTAR} image
of the field containing J0811. The color scale is arbitrarily adjusted
for better visibility. {\it Fermi}/LAT 3FGL ellipse (95\%, white) and the best-fit circle (95\%, magenta)
are shown, and a $R=30''$ circle is drawn around the serendipitous source (denoted as J0810.0$-$7527).
{\it Middle}: Location of the sources we are studying in the {\it WISE}
[3.4]$-$[4.6]$-$[12]$\mu m$ color-color diagram \citep[Figure taken from][]{dma+12}.
The four sources, including J0810, are located in the middle of the BZB
(naming convention for BL Lac in the ROMA-BZCAT catalog) distribution.
See \citet{dma+12} for more detail.
{\it Right}: Observed SED of the serendipitous source. Note that we used
$N_{\rm H}=6.9\times 10^{20}\rm \ cm^{-2}$, the optical extinction inferred value,
for constructing the SED. Notice that this new source is quite
hard, emitting more strongly in the {\it NuSTAR} band than in the {\it XMM-Newton} band.
\label{fig:fig2}
}
\vspace{0mm}
\end{figure*}

We discovered a serendipitous X-ray source (J0810) in the field of J0811 (Figure~\ref{fig:fig2}).
The X-ray ({\it XMM-Newton}) position of the source is
R.A. = 08\textsuperscript{h}10\textsuperscript{m}03\textsuperscript{s}
and decl. = $-$75$^\circ$27$'$21$''$ (J2000, $\delta_{\rm R.A.,\ decl.} =2''$ statistical only),
only 6$'$ from J0811 (Figure~\ref{fig:fig2} left).
We find that the spectrum cannot be described with a simple absorbed power law
($\chi^2$/dof=185/118, $p=7\times10^{-5}$).
A broken power-law model\footnote{http://heasarc.gsfc.nasa.gov/docs/xanadu/xspec/manual/XS\\modelBknpower.html}
explains the data ($\chi^2$/dof=116/116, $p=0.47$) and
the best-fit parameters are $N_{\rm H}=1.4\pm0.3\times10^{21}\rm \ cm^{-2}$,
low-energy photon index $\Gamma_{\rm 1}=3.4\pm0.3$, high-energy photon index
$\Gamma_{\rm 2}=1.74\pm0.07$, break energy $E_{\rm break}=1.46\pm0.08$\,keV
and 3--10\,keV flux $F_{\rm 3-10 keV}=2.7\pm0.2\times 10^{-13}\rm \ erg \ s^{-1}\ cm^{-2}$.

Together with archival radio, optical, and {\it Swift} UV data, we
construct the SED of the source (Figure~\ref{fig:fig2} right). If we use the best-fit X-ray
$N_{\rm H}$, the extrapolated spectrum matches poorly to the optical. Instead
we de-absorb using the value from the optical/UV extinction $N_{\rm H}=6.9\times10^{20}\rm \ cm^{-2}$.
X-ray fits with absorption fixed at this value are statistically acceptable
(null hypothesis probability $p=0.3$).
The SED of this source suggests a blazar with $\nu^{\rm sy}_{\rm pk}$ in the optical range,
and a rise to a Compton component in the hard X-ray band. Its location in the
{\it WISE} color-color diagrams \citep[Figure~\ref{fig:fig2} middle; see also][]{dma+12}
suggests that the source should be a BL Lac.
If the Compton component peaks at $>100$\,MeV, this source may contribute to the J0811 SED,
since the source is within the aperture we used for J0811. If we free the position of J0811 in the
{\it Fermi} analysis, we find a maximum likelihood coincident with J0811
(magenta circle in Figure~\ref{fig:fig2} left). Also, a second source
at the J0810 position does not significantly increase the model test statistic (TS).

We then increased the zenith angle cut to $<100^\circ$ to have more events
and used a small spatial bin size (0.05$^\circ$) to see if J0810 is detected
in the {\it Fermi} band. We performed binned likelihood analysis with
the new data. In this case, a gamma-ray counterpart
of J0810 is detected significantly (TS=56); the model without J0810 is only 0.03\% as
probable as the one with J0810.
In the 0.1--500\,GeV band, J0810 has $\sim$20\% of the flux (with 40\% flux uncertainty)
of J0811 with a similar power-law index ($\Gamma_{\gamma}=1.8\pm0.1$).
These spectral parameters for J0810 may not be very accurate because of mixing from
the brighter source, J0811.
Since J0811 is brighter than J0810 in the gamma-ray band,
we attribute all of the LAT flux to J0811 in SED modeling
and discuss implication of J0810 contamination on the model (see Section~\ref{sec:sec3_3}).

\begin{figure*}
\centering
\vspace{-75.0 mm}
\hspace{-12.0 mm}
\includegraphics[width=5.7 in,angle=90]{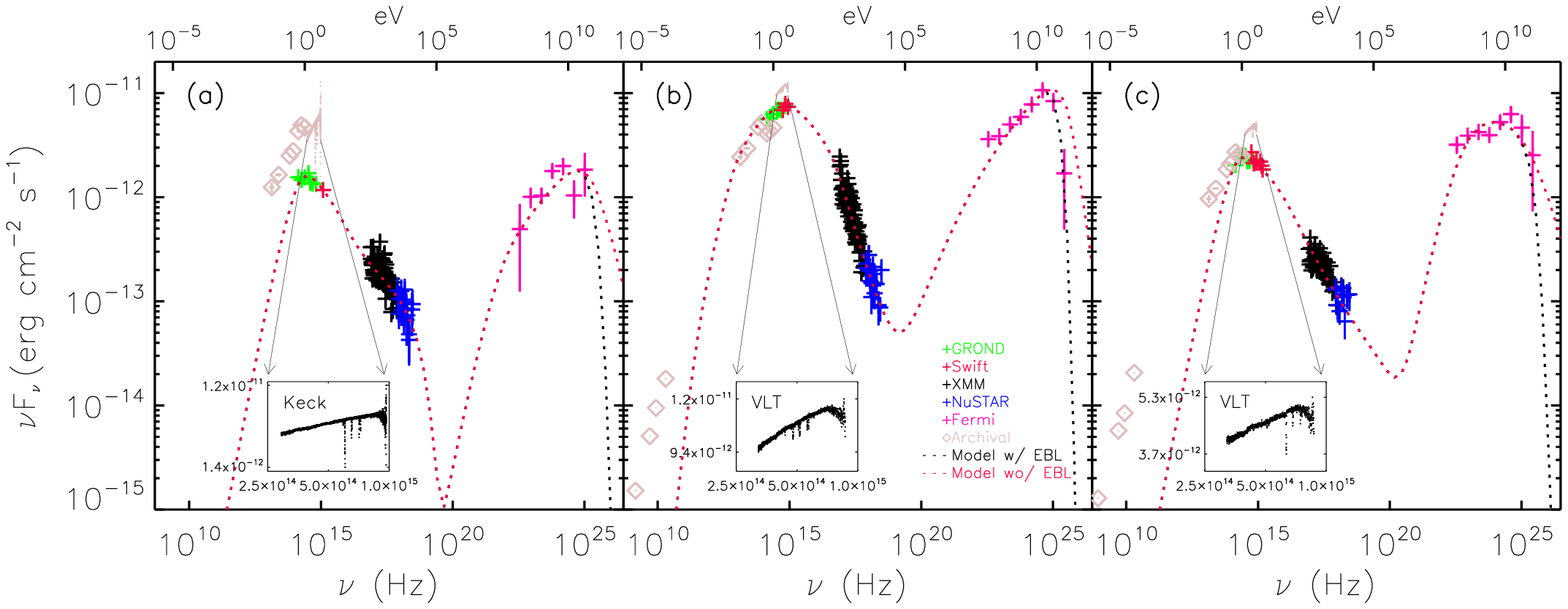} \\
\vspace{-5.0 mm}
\figcaption{Observed broadband SED and best-fit models for
({\it a}) J0022, ({\it b}) J0630, and ({\it c}) J0811.
Data points with an error bar are taken from the contemporaneous
observations (Sections~\ref{sec:sec2_1} and \ref{sec:sec2_2})
and diamonds are from the archival observations (Section~\ref{sec:sec2_3}).
The dashed lines are the best-fit SSC SED models
of \citet{bms97} with (black) and without (red) EBL absorption \citep[][]{frd10}.
Note that the archival data are not taken contemporaneously even if they
are plotted in the same color and symbol. The insets plot the VLT/Keck spectra of
\citet{src+13}, showing the lack of emission lines and the low excitation
absorption complexes placing lower limits on the redshift. These observations
appear to have been in a brighter, harder optical state.
\vspace{2.0 mm}
\label{fig:fig3}
}
\vspace{0mm}
\end{figure*}

\section{Data Analysis and Results}
\label{sec:sec3}

\subsection{Variability}
\label{sec:sec3_1}

        We have examined the collected data for variability, since short
timescales can give useful constraints on the characteristic size of the emission zone
in the various wavebands. We first examined our contemporaneous data sets for
short-timescale variations. For the {\it XMM-Newton} and {\it NuSTAR} data, spanning
$\sim$10--100\,ks, we constructed exposure-weighted light curves using various time bin sizes
($\sim$100--20,000~s), ensuring $> 20$ counts in each time bin, and calculated $\chi^2$
for a constant flux. The probability for constancy was always high ($\gapp$10\%),
implying no significant short-term variability for any of the three sources at this epoch.
Similarly, the optical/UV data from the contemporaneous epoch did not show sub-day variability.

        However, on longer time scales, the optical synchrotron peak flux does show
substantial variability, as can be seen by comparing the contemporaneous and archival points
in Figures~\ref{fig:fig1} and \ref{fig:fig3}. J0022, for example varies by $\sim 6\times$.
As noted, the VLT/Keck spectra also appear to represent brighter epochs, although slit losses limit
the precision of the flux calibration. In general, the brighter epochs appear to have harder
near-IR to UV spectra, suggesting increased electron energy (or increased bulk Doppler factor)
in flaring events. A much better characterization of J0630's optical variability is available from
the KAIT {\it Fermi} AGN monitoring data \citep[][]{crf+14}.\footnote{http://brando.astro.berkeley.edu/kait/agn/}
The dominant modulation is slow on $\sim$year timescales; this is of modest amplitude compared
to other BL Lacs ($\sim$50\%). KAIT resolves times as short as the $\sim$3d cadence and
we do see statistically significant ($\gapp$6$\sigma$) changes between consecutive observations.
This suggests that at least some of the jet flux arises in compact $r<10^{16}$cm structures.

        We can use the LAT band to probe variability in the Compton peak emission. Since these
sources are not very bright, we were able to only probe $\sim$month timescales. To
this end we generated lightcurves by fitting source fluxes to 100\,MeV--500\,GeV
photons from a 5 degree ROI about each source using the {\tt gtlike} tool for each time bin.
For this we fixed the background model
normalization and the background source spectral parameters at the mission-averaged values (see below),
allowing only the source flux to vary with the spectral index held fixed at the values given
in Table~\ref{ta:ta2} .
Figure~\ref{fig:fig1} shows the corresponding light curves. The variability
is not strong ($\chi^2$/dof values for a constant light curve of 
5/8, 92/72, and 28/33 for J0022, J0630 and J0811, respectively).
We confirm the results of the 3FGL catalog \citep[][]{fermi3fgl};
our sources are not flagged as variable in the 3FGL catalog at a 99\% confidence.
Finally, examination of light curves assembled by the
Agenzia Spaziale Italiana science data center\footnote{http://www.asdc.asi.it/fermi3fgl/}
also shows no significant variability in any source. We conclude that the three sources have
been relatively quiescent for BL Lacs -- this gives us confidence that the mission-averaged LAT
spectrum may be usefully compared with our contemporaneous campaign fluxes for SED fitting.

\subsection{Constructing broadband SEDs}
\label{sec:sec3_2}

\newcommand{\markaa}{\tablenotemark{a}}
\newcommand{\markbb}{\tablenotemark{b}}
\newcommand{\markcc}{\tablenotemark{c}}
\begin{table}[t]
\vspace{-0.0in}
\begin{center}
\caption{Galactic foreground reddening values and X-ray/gamma-ray fit results
\label{ta:ta2}}
\vspace{-0.05in}
\scriptsize{
\begin{tabular}{cccccccc} \hline\hline
Source			&				& J0022   & J0630   & J0811 \\ \hline
$E(B-V)$		& (mag)				& 0.024   & 0.056   & 0.125 \\
$N_{\rm H,\ Dust}$\markaa	& ($10^{20}\rm \ cm^{-2}$)	& 1.3     & 3.1     & 7      \\
$N_{\rm H}$		& ($10^{20}\rm \ cm^{-2}$)	& 4(1)    & 13(1)   & 7(1)   \\
$\Gamma_{\rm X}$	& $\cdots$			& 2.55(6) & 2.98(7) & 2.45(7) \\
$F_{\rm X}$\markbb	& $\cdots$			& 0.93(8) & 1.6(1)  & 1.4(1) \\
$\Gamma_{\rm \gamma}$	& $\cdots$			& 1.86(6) & 1.83(3) & 1.93(4) \\
$F_{\rm \gamma}$\markcc	& $\cdots$			& 6.3(9)  & 25(2)   & 23(2) \\ \hline
\end{tabular}}
\end{center}
\vspace{-0.5 mm}
$^{\rm a}$ Dust-extinction equivalent $N_H$, converted with
$N_{\rm H}=1.8\times 10^{21} A(V)\rm \ cm^{-2}\ mag^{-1}$ and $R_V=3.1$ \citep{ps95}.\\
$^{\rm b}$ 3--10\,keV flux in units of $10^{-13}\rm \ erg\ s^{-1}\ cm^{-2}$.\\
$^{\rm c}$ 0.1--500\,GeV flux in units of $10^{-9}\rm \ photons \ s^{-1}\ cm^{-2}$.
\end{table}

Next we assembled broadband SEDs for the sources using the data described in Section~\ref{sec:sec2}.
The optical/UV magnitudes were corrected for the dust map extinction in these directions
(Table~\ref{ta:ta2}) obtained from the NASA/IPAC extragalactic database, using
the \citet{sf11} calibration. We show the SEDs in Figure~\ref{fig:fig3}.
Note that Lyman-$\alpha$ forest absorption was visible in J0630 at frequencies above
$\sim10^{15}$\,Hz in the UVOT data, as expected from its large redshift;
we do not use the high-frequency UVOT data $\gapp 10^{15}$\,Hz in the J0630 SED modeling.

The X-ray response files are produced with the standard tools in SAS and in {\tt nustardas}
for the {\it XMM-Newton} and {\it NuSTAR} spectra, respectively.
We fit the spectra in the 0.3--79\,keV band with an absorbed power-law model
in {\tt XSPEC} 12.8.2 and found that the model describes the data well, having
$\chi^2$/dof$\lapp$1 for all three sources. The fact that all X-ray spectra are
well modeled by a single absorbed power law is important to the modeling below.
The absorption corrections for the X-ray data were obtained from the $N_{\rm H}$
in the power-law fits.  The fit results are presented in Table~\ref{ta:ta2}.

        While the X-ray fit and extinction map values for the absorption agree well for J0811,
J0022 and especially J0630 show stronger X-ray absorption. Given the modest dust map resolution,
and the $\sim$50\% conversion uncertainties \citep[e.g.,][]{g75,w11,fgos15}, the discrepancy
for J0022 may be reconciled. However the large value for J0630 seems difficult to accommodate
and we have no clear explanation. The Galactic H{\scriptsize I} column
density\footnote{https://heasarc.gsfc.nasa.gov/cgi-bin/Tools/w3nh/w3nh.pl}
toward J0630 is 7--12$\times 10^{20}\rm cm^{-2}$, consistent with the X-ray inferred value.
If we assume the X-ray value for de-extinction of the optical, we find an unnatural UV flux rise
(similarly, using the optical value makes an unnatural cutoff in the low energy X-ray spectrum).
Thus we can only accommodate the X-ray fit value if the optical/UV flux has an extra
blue, narrow-band component. This seems unnatural. Alternatively the dust map extinction might
be correct and the X-ray component may be spatially separated from the optical emission,
experiencing extra local (host) absorption.
Measuring the J0630 VLT absorption line strengths indicated that the intervening/host
galaxy supplies negligible extinction $E(B-V)<0.01$ to the optical component, which is
consistent with the low effective $E(B-V)$.
Acknowledging this inconsistency, we use the two values in Table~\ref{ta:ta2}
when constructing the SED.

For the {\it Fermi} SED, we performed binned likelihood analysis
using the same configuration as described in Section~\ref{sec:sec2_2} with the 6.5-yr data.
In doing so, we fit spectra for all bright sources (detected with $\gapp5\sigma$) in the ROI
and the background amplitudes. Spectral parameters for faint sources or those outside the ROI
are held fixed at the 3FGL values.
The results are shown in Table~\ref{ta:ta2}.
The highest-energy bands in which a significant detection (TS$>15$)
was made are 29--75\,GeV, 75--194\,GeV, and 75--194\,GeV for J0022,
J0630 and J0811, respectively (see Figure~\ref{fig:fig3}).
We then derive the SEDs using the best-fit power-law model, and show the inferred spectrum in
Figure~\ref{fig:fig3}, where the TS is greater than 15
for each data point.  We performed the analysis using different ROI sizes,
finding consistent results.
In Figure~\ref{fig:fig3} we show the results obtained for the 5$^\circ$ extraction
as it gives the highest TS value.

We show the broadband SEDs in Figure~\ref{fig:fig3}.
A non-contemporaneous broadband SED for J0630 with sparser X-ray and gamma-ray data
has been previously reported \citep[][]{gtf+12, pgr12}; the results are broadly similar to
our measurements.

\subsection{SED modeling}
\label{sec:sec3_3}

\newcommand{\markax}{\tablenotemark{a}}
\newcommand{\markbx}{\tablenotemark{b}}
\begin{table*}[]
\vspace{-0.0in}
\begin{center}
\caption{Best-fit parameters for the SSC model of B97 with single power-law injection
\label{ta:ta3}}
\vspace{-0.05in}
\scriptsize{
\begin{tabular}{ccccc} \hline\hline
Parameter           & Symbol     & 3FGL~J0022.1$-$1855 & 3FGL~J0630.9$-$2406 & 3FGL~J0811.2$-$7529 \\ \hline
Redshift            & $z$        & 0.774        & $>1.239$  & 0.689 \\
Doppler factor      & $\delta_{\rm D}$ & 19    & 71     & 33 \\
Bulk Lorentz factor & $\Gamma$   & $>9.6$   & $>35.3$  & $>16.5$  \\
Viewing angle (deg.) & $\theta_v$ & $<3.0$ & $<0.81$  & $<1.74$ \\
Magnetic field (mG) & $B$        & 60        & 1016      & 7 \\
Comoving radius of blob (cm) & $R'_b$ & $1.12\times10^{14}$    & $1.78\times10^{13}$  & $1.52\times10^{14}$  \\ 
Effective radius of the blob ($cm$) & $R'_{\rm E}=(3 R'^{2}_b t_{\rm evol} c/4)^{1/3}$ & $1.4\times10^{15}$ & $1.9\times10^{14}$  & $1.7\times10^{15}$ \\ \hline
Initial electron spectral index & $p_{\rm 1}$ & 3.14       & 4.26   & 3.19        \\ 
Initial minimum electron Lorentz factor & $\gamma'_{\rm min}$ & $2.88\times 10^{4}$ & $1.41\times 10^4$  & $1.18\times 10^{4}$ \\
Initial maximum electron Lorentz factor & $\gamma'_{\rm max}$ & $1.5\times10^{6}$ & $2.7\times10^7$ & $3\times10^{7}$ \\
Injected particle luminosity (erg s$^{-1}$)\markax & $L_{\rm inj}$ & $9\times10^{42}$ & $7\times10^{41}$  & $8\times10^{42}$ \\ 
$\chi^2$/dof & $\cdots$ & 151.1/122 & 186/140 & 128.5/94 \\ \hline
Synchrotron peak frequency (Hz)\markbx & $\nu^{\rm sy}_{\rm pk}$   & $5.6\times10^{14}$  &  $1.5\times10^{15}$ &  $5.8\times10^{14}$     \\
Synchrotron peak luminosity($\rm erg\ s^{-1}$)\markbx & $L^{\rm sy}_{\rm pk}$ & $4.6\times10^{45}$  & $6.7\times10^{46}$  &  $5.1\times10^{45}$   \\
Compton dominance & CD    & 1.2  & 1.4  & 2.1  \\ \hline \hline
\end{tabular}}
\end{center}
\hspace{-2.0 mm}
$^{\rm a}${Energy injected into the jet in the jet rest frame \citep[see][]{bc02}.}\\
$^{\rm b}${Quantities in the observer frame.
The luminosity quoted is that inferred assuming isotropic emission.}\\
\end{table*}

         We use the one-zone synchro-Compton model of \citet[][hereafter B97]{bms97} to model
the SEDs of the sources. The code evolves a spherical blob of electron/positron
plasma with a power-law injected energy distribution,
following the e$^+$/e$^-$ population over $10^7$\,s ($t_{\rm evol}$)
assuming that the particle energy loss is dominated by radiative cooling
as the blob zone flows along a jet axis.
As blobs are continuously injected, the emission zone forms
a cylindrical shape (i.e., jet)
elongated along the jet axis ($l=ct_{\rm evol}=3\times10^{17}$\,cm)
and the time-integrated spectrum determines the jet emission.
The effect of pair-absorption is calculated and included in the model.
The full model has 16 parameters including those for disk and BLR emission;
to simplify we start with standard BL Lac assumption
that self-Compton emission dominates so that the seed photons from BLR
and disk are negligible.  The seven remaining parameters we adjust are
the low-energy and high-energy cutoffs ($\gamma'_{\rm min,max}$)
and spectral index of the power-law electron distribution ($p_{\rm 1}$),
the magnetic field strength ($B$), the bulk Lorentz factor of the jet ($\Gamma$)
(this is done for a fixed viewing angle $\theta_{\rm v}$,
hence equivalent to adjusting the Doppler factor $\delta_{\rm D}$)
and the blob rest frame size ($R'_b$) and electron density ($n_{e}$),
which serve to normalize the total flux.
This model has also been used for modeling SED of other blazars \citep[e.g.,][]{hba+01,r06}.

         We use the following steps to find best-fit SED parameters:
(1) adjust the parameters to visually match the SED for initial values,
(2) vary each individual parameter over a range
(a factor of $\sim$2 initially and decreased with iterations) with
ten grid points while holding the other parameters fixed,
(3) find the parameter value that provides the minimum $\chi^2$,
(4) update the parameter found in step (3) with the best-fit value,
(5) repeat (2)--(4) until the fit does not improve any more.
Because the X-ray spectra are so well described by a simple power law,
we initially identify their spectra with synchrotron emission of a cooled electron population,
strongly constraining the fit parameter set.
We do not include the highest energies ($\gapp 40$\,GeV) LAT points in the initial
fits, as we will use them later for EBL constraints as done by \citet{dfp+13}. We update
only one parameter each iteration although we vary all seven parameters.
We present the best-fit parameters in Table~\ref{ta:ta3}.
We also measured $\nu^{\rm sy}_{\rm pk}$, $L^{\rm sy}_{\rm pk}$ and CD
using the best-fit SED model, and present them in Table~\ref{ta:ta3}.

        In the model $\Gamma$ and $\theta_{\rm v}$
appear only in combination through the Doppler factor $\delta_{\rm D} = [\Gamma(1-\beta\mu)]^{-1}$,
where $\beta=\sqrt{1-1/\Gamma^2}$ and $\mu=\mathrm{cos}(\theta_{\rm v})$. Hence, the model determines
only $\delta_{\rm D}$ unless one has external constraints on one of $\Gamma$ or $\theta_{\rm v}$.
Therefore, for a given $\delta_{\rm D}$, only lower and upper limit for $\Gamma$ and $\theta_{\rm v}$ can
be inferred, also given in Table~\ref{ta:ta3}.

        While the procedure above converges well to a local minimum, there is always a risk
that quite distinct solutions could provide better fits. The high dimensionality of the fit space,
plus the incomplete SED coverage makes it difficult to locate such minima. To aide our exploration of
parameter space, we used the initial scans to define the covariance between the various
quantities. We find that simple power-law co-dependencies capture most of the covariance
trend around the fit minimum. We fit an amplitude and slope for each parameter pair. Thus,
by varying one
control parameter, say $B$, and then setting the others to the covariance-predicted values, we can
take larger steps without wandering too far from the $\chi^2$ minimum surface. For each
such trial solution, we then compute small test grids to rapidly converge to the local
minimum (with the control parameter held fixed). In this way we explored the minima
connected to the `best fit' solution tabulated above. This gave us larger ranges for
`acceptable' (i.e.  null hypothesis probability $p>0.01$) solutions. For example for J0630
acceptable solutions were found for $0.3$\,G$<B<3$\,G, although all were poorer fits than the
best solution (Table~\ref{ta:ta3}).

       We note that J0811 flux in the {\it Fermi} band may be lower by $\sim$20\% than is used
in the modeling if we remove J0810 contamination (see Section~\ref{sec:sec2_4}).
We therefore performed {\it Fermi} data analysis including J0810 and constructed
a new SED of J0811. We modeled the new SED as described above and found that
significant changes need to be made only for parameters related to
high-energy normalization, and our conclusion on EBL constraints below remains the same.

\subsection{EBL Constraints}
\label{sec:sec3_4}

        We have been careful not to use the highest energy LAT points in the SSC SED
fits, although we see that all models over-predict the high-energy LAT
flux.  We now apply EBL models to the data and calculate $\chi^2$ with and without
EBL models, showing the results in Table~\ref{ta:ta4} (see also Figures~\ref{fig:fig3}--\ref{fig:fig5}).
Note that we used all the SED data including those $>40$\,GeV here.
Not unexpectedly, EBL absorption provides no significant improvement to the fits
of the lower redshift sources J0022 and J0811.  However, we see clear
improvements ($\Delta\chi^2\sim10$ corresponding to $\sim5\sigma$)
for J0630. Only the high UV model provides no improvement.
The $\chi^2$ decrease is similar for the more conventional models.

\begin{table}
\vspace{-0.0in}
\begin{center}
\caption{Best-fit $\chi^2$ values for the EBL models tested in this work
\label{ta:ta4}}
\vspace{-0.05in}
\scriptsize{
\begin{tabular}{cccccccc} \hline\hline
Model                 & J0022  & J0630 & J0811 & reference \\  \hline
No EBL                & 151.1  & 197.4 & 128.9 & $\cdots$  \\
Dom{\'{\i}}nguez      & 151.1  & 186.2 & 129.6 & [1]  \\
Franceshini           & 151.1  & 186.2 & 129.6 & [2]  \\
Gilmore Fiducial      & 151.0  & 189.2 & 130.0 & [3]  \\
Gilmore Fixed         & 151.1  & 186.5 & 129.6 & [3]  \\
Helgason              & 151.1  & 186.3 & 129.5 & [4]  \\
Kneiske04 best fit    & 151.1  & 191.4 & 130.6 & [5]  \\
Kneiske \& Dole       & 151.1  & 187.4 & 129.8 & [6]  \\
Kneiske high UV       & 150.3  & 205.1 & 132.8 & [5]   \\
Stecker high opac.    & 151.0  & 194.0 & 131.6 & [7]  \\
Stecker low opac.     & 151.0  & 187.4 & 130.2 & [7]  \\
Finke `C'             & 151.1  & 187.0 & 129.7 & [8]  \\ \hline
\end{tabular}}
\end{center}
\footnotesize{
References: [1] \citet{dpr+11}
[2] \citet{frv08}
[3] \citet{gsp+12}
[4] \citet{hk12}
[5] \citet{kbm+04}
[6] \citet{kd10}
[7] \citet{sms12}
[8] \citet{frd10}
}\\
\end{table}

        Since the redshift measurement for J0630 is only a lower limit,
we attempted to fit $z$ in the EBL model fits.
Allowing one more free parameter (holding the other parameters fixed)
improves the fit in general but the improvement
is small except for the case of the disfavored models.
For all models the best-fit $z$ is less than the spectroscopic lower limit, although
this is within errors for the best-fit models. Accordingly, we hold $z$ fixed at 1.239.

        Although the LAT observations continue, unless there is a strong flare,
we are unlikely to greatly improve the J0630 EBL constraints without going to higher
energy. This will be challenging with present and future generation air Cerenkov telescopes;
we predict an absorbed 200\,GeV energy flux
of $\nu F_{\nu} \sim4\times10^{-14}\rm \ erg\ cm^{-2}\ s^{-1}$ which is
an order of magnitude lower than the 5-$\sigma$ sensitivity of the Cherenkov
Telescope Array\footnote{https://portal.cta-observatory.org/Pages/Home.aspx}.
Further LAT study of other high-redshift BL Lacs
can certainly probe the EBL evolution at $z>1.5$.

\subsection{Alternative Fits}
\label{sec:sec3_5}

\begin{figure}
\centering
\hspace{-5.0 mm}
\includegraphics[width=3.5 in]{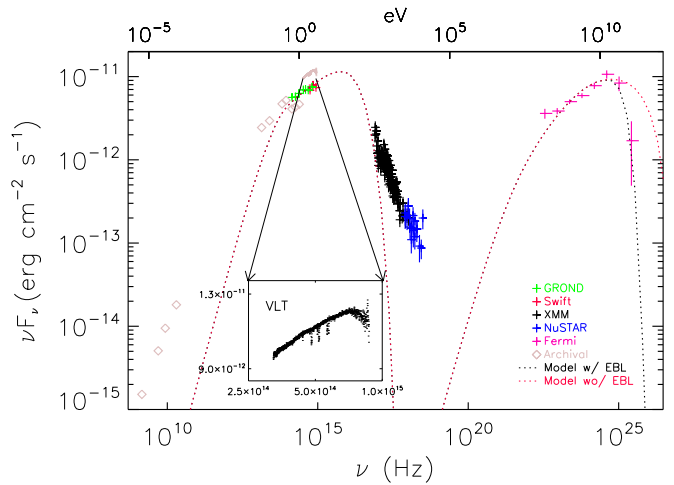} \\
\figcaption{An SED model fit with (black dotted line) and without (red dotted line)
the EBL absorption model (Finke `C' in Table~\ref{ta:ta4}) for the J0630 data
with a hard injection spectrum. The parameters
for this model are:
$\delta_{\rm D}=73$,
$\theta_{\rm v}=0.74^\circ$, $B=10$\,mG,
$R'_{\rm b}=2\times10^{14}\rm cm$, $p_{\rm 1}=2.35$,
$\gamma_{\rm min}=5\times 10^{3}$,
and $\gamma_{\rm max}=2\times10^5$.
\label{fig:fig4}
\hspace{-10mm}
}
\end{figure}

The best-fit parameters for our BL Lacs are unusual with steep $p_{\rm 1}>3$ injection spectra.
J0630 is the most extreme, with $p_{\rm 1}\approx 4.3$ and a strong $\sim 1$\,G magnetic field.
The excellent power-law fits to the {\it XMM-NuSTAR} X-ray data drive these values.
We have attempted to fit J0630 with more conventional $2<p_{\rm 1}<3$ indices, but such models
are always strongly excluded by the X-ray spectral points. The only option is to remove the X-ray
points from the fits, assign them to an additional, unmodeled component. Then excellent fits
to the rest of the SED with more conventional, lower $p_{\rm 1}$ and $B$ values can be obtained,
an example of which
is shown in Figure~\ref{fig:fig4}. The synchrotron peak energy is higher (consonant with
the high source power) and the X-rays are under-predicted; the observed spectrum is an additional,
soft component.
This soft component, if produced by synchrotron emission, can be generated by an electron distribution
with $\gamma_{\rm min}'>4\times10^{4}$, $\gamma_{\rm max}'=5\times10^{6}$, $p_{\rm 1}=4.1$
and a small electron density $\sim 10^{-1}\rm cm^{-3}$ in order not to
overproduce the optical and the Compton emission.

        We are focused on the LAT band fit, so it is interesting to see that
this model has a very similar cutoff to that of Figure~\ref{fig:fig3}b, requiring a similar EBL absorption.
The $\chi^2$ values (18 data points ignoring the X-ray data) are 62 and 86 with and without the
EBL absorption, respectively.
Evidently inverse Compton emission from the X-ray component, if any, is in the highly absorbed
TeV band. We can speculate that the soft X-ray component rises in
a different zone of the jet \citep[e.g.,][]{m14},
arguably with large $B$ and a steep, highly cooled spectrum. Whether this connects to the apparently
different absorption for this component is unclear.

\begin{table*}[t]
\vspace{-0.0in}
\begin{center}
\caption{Best-fit parameters of the FDB08 model
\label{ta:ta6}}
\vspace{-0.05in}
\scriptsize{
\begin{tabular}{ccccc} \hline\hline
Parameter           & Symbol     & J0022.1$-$1855 & 3FGL~J0630.9$-$2406 & 3FGL~J0811.2$-$7529 \\ \hline
Redshift            & $z$        & 0.774        & $>$1.239        & 0.689        \\
Doppler factor      & $\delta_{\rm D}$ & 29      & 110     & 49       \\
Magnetic field (mG) & $B$        & 37           & 4.7          & 7.9          \\
Variability timescale (s) & $t_v$ & $10^5$      & $10^5$       & $10^5$        \\
Comoving radius of blob (cm) & $R'_b$ & $4.9\times10^{16}$    & $1.5\times10^{17}$  & $8.7\times10^{16}$  \\ \hline
Lower-energy electron spectral index & $p_{\rm 1}$ & 2.5      & 2.4      & 2.6        \\
High-energy electron spectral index & $p_{\rm 2}$ & 4.0      & 4.5      & 4.0        \\
Minimum electron Lorentz factor & $\gamma'_{\rm min}$ & $6\times10^{3}$ & $10^3$  & $3\times10^{3}$ \\
Break electron Lorentz factor & $\gamma'_{brk}$ & $3.9\times10^{4}$ & $6.9\times10^4$ & $4.9\times10^{4}$ \\
Maximum electron Lorentz factor & $\gamma'_{\rm max}$ & $3.0\times10^{6}$ & $3.0\times10^6$  & $6\times10^{6}$ \\ \hline
\end{tabular}}
\end{center}
\hspace{-2.0 mm}
\vspace{2.0 mm}
\end{table*}

\begin{figure*}
\centering
\vspace{-80.0 mm}
\hspace{-12.0 mm}
\includegraphics[width=5.7 in,angle=90]{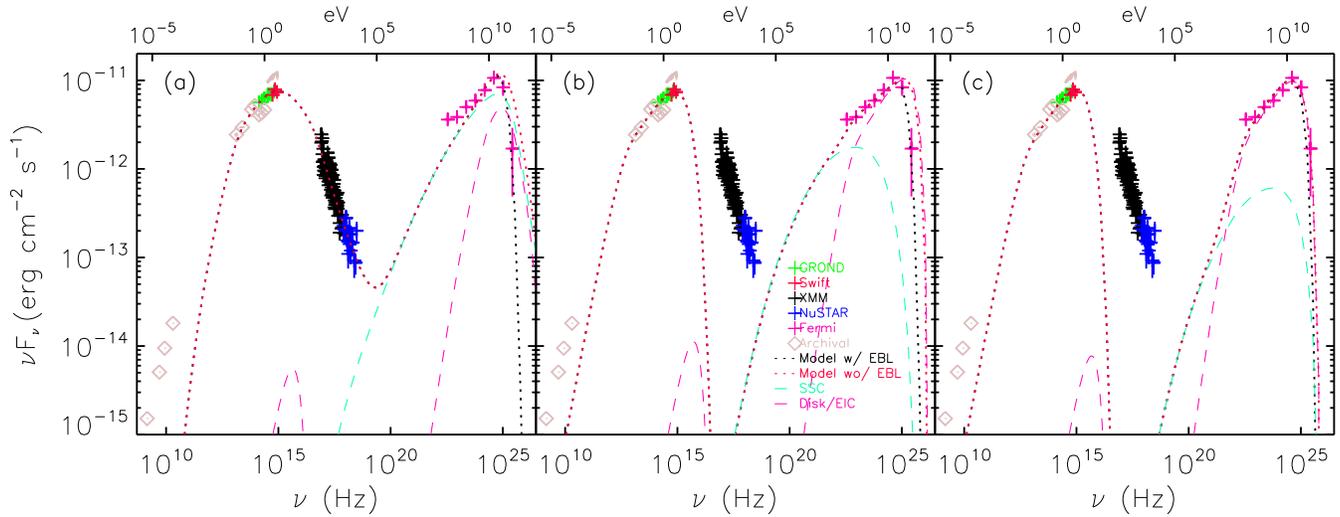} \\
\vspace{-4.0 mm}
\figcaption{SED models with the disk component for J0630.
({\it a}): A model with the disk component added to the baseline synchrotron+SSC model
in Figure~\ref{fig:fig3}b.
({\it b}): Similar to (a), but baseline model is that in Figure~\ref{fig:fig4}.
({\it c}): Same as (b) with larger $B$ and lower $\gamma_{\rm max}$.
The model parameters are further adjusted from the baseline ones to match the SED.
The EBL model we used for the plot is the ``Finke C'' model in Table~\ref{ta:ta4}.
See text for more details.
\label{fig:fig5}
}
\end{figure*}

If we allow an additional X-ray emitting component, we might also consider a more
complex injection model \citep[][hereafter FDB08]{fdb08}.
We try an electron distribution that is a broken power law or a log parabola.
To compare parameters, we fit to this model by first choosing
a variability timescale and then adjusting the other parameters ($\delta_{\rm D}$, $B$, 
and the electron distribution) until a good fit was obtained.
We assumed $t_v=10^{5}$\,s which is consistent with the timescale for the optical flux variability
in J0630 ($t_v\lapp3$\,days). The broken power-law model is always more satisfactory than
the log-parabola version and we show the best-fit parameters for our three BL Lacs in Table~\ref{ta:ta6}.
It is interesting to compare to our cooling model fits. In particular, the power law breaks strongly
to large $p_{\rm 2}$ values. This is imposed by fiat here, but the drive to such large break
is difficult to accommodate in self-consistent cooling and can require large
magnetic field strengths (Table~\ref{ta:ta3}).
We conclude that if conventional $p_{\rm 1}\sim2-3$ electron injection spectra
are adopted, we will always require an additional steep component not easily achieved by
radiative cooling.

        We have noted that the $>$GeV LAT spectrum is not affected by this extra electron component
(and thus our EBL conclusions for J0630 are robust). However this is in the context of SSC models.
\citet{gtf+12} and \citet{pgr12} noted that HSP BL Lacs
can also have low level disk/BLR emission, overwhelmed by
(and   invisible behind) the jet synchrotron component along the Earth line-of-sight, yet
providing substantial seed photons for Compton up-scatter. These may have significant impact on
the high-energy hump of the SED \citep[blue FSRQ model;][]{gtf+12, pgr12}.
Thus, we explore  B97 model for J0630
with a disk component (orders of magnitude fainter than the baseline synchrotron emission)
which can produce additional Compton emission at $\sim 10^{24}-10^{26}$\,Hz (Figure~\ref{fig:fig5}).
We assume a small BL covering fraction given the strong limits
on broad line equivalent widths \citep[][]{src+13}.

        In Figure~\ref{fig:fig5}a, we add disk EC emission to the model of
Figure~\ref{fig:fig3} with a soft ($p_{\rm 1}=4.26$) injection spectrum. The strong constraint
of the X-ray data preclude any large change in the SSC component. We find that the additional
EC emission contributes primarily at high LAT energies.  The net effect is to under-produce
the low energy gamma-rays leading to an excessively hard LAT spectrum, while not significantly
changing the high-energy spectral shape. Thus the EC is not statistically demanded by  this model,
but even if EC is added, significant EBL absorption should be present;
improvement of the fit when the EBL models in Table~\ref{ta:ta4} are included is typically
$\Delta \chi^2\sim20$.

Addition of the disk/EC component to the model in Figure~\ref{fig:fig4}
(hard injection spectrum) provides more flexibility since we do not need to match the
X-ray spectrum, having assumed above that the X-ray emission in this model
is from a different region than the peak jet emission.
In this case, the shape of the SSC component can be adjusted to match the low-energy
gamma-ray data and the EC emission accounts for the higher energy data (Figure~\ref{fig:fig5}b);
this model reproduces
the optical/UV and gamma-ray data better than the baseline model (Figure~\ref{fig:fig4})
does. Nevertheless, the effect of EBL absorption is clearly visible in Figure~\ref{fig:fig5}b,
and including the EBL models improves the fit by $\Delta\chi^2\sim40$.

It may be imagined that the sharp drop above $10^{25}$\,Hz in the unabsorbed model
(dashed magenta line in Figure~\ref{fig:fig5}b) may be able to reproduce
the sharp drop in the SED without a visible effect of the EBL absorption
if the peak frequency of the EC component can be lowered.
This can be done by lowering $\gamma_{\rm max}'$, but merely adjusting
$\gamma_{\rm max}'$ will damage the goodness of fit in the optical-UV band.
However, by adjusting $B$, $\gamma'_{\rm max}$, and $\Gamma$ ($\delta_{\rm D}$),
lowering only $\nu^{\rm IC}_{\rm pk}$ without affecting
$\nu^{\rm sy}_{\rm pk}$ is possible since the latter is $\propto \Gamma B\gamma_{\rm max}'^2$
while the former is $\propto \Gamma^2 \gamma_{\rm max}'^2$.
We first adjust $B$ (decrease) and $\gamma_{\rm max}'$ (increase), and find that
$\nu^{\rm sy}_{\rm pk}$ is also lowered in this case owing to stronger cooling caused
by the stronger magnetic field strength. So we lower $\Gamma$, and adjusted $B$ and $\gamma_{\rm max}'$.
In this way, we were able to match the steep fall in the SED at $\gapp 10^{25}$\,GHz
without invoking EBL absorption (Figure~\ref{fig:fig5}c). For this model, we use
$B=15$\,mG, $\gamma_{\rm max}'=8\times 10^{4}$ and $\delta_{\rm D}=27$
(corresponding to $\Gamma>14$ and $\theta_v<2.1^\circ$).
In this case, as we intended, the fit
is better when the EBL absorption is not considered; the EBL effect makes
the model underpredict the data, and including the EBL models increases
$\chi^2$ by $\sim3$ typically.
Note that for models in Figures~\ref{fig:fig5}b and c,
we assumed that there is a sharp high-energy cutoff in the synchrotron emission.
However, if such a sharp cutoff does not exist,
the high-frequency SSC/EC component should be enhanced, perhaps similar to that
in figure~\ref{fig:fig5}a, requiring the EBL absorption.

Note that we can also add BLR-reflected disk photons to this
model \citep[see][for example]{r06}. The EC emission of the reflected photons
only appears at higher frequencies than the direct disk component and thus
suffers from severe EBL absorption. Therefore, we do not consider this component here.

\section{Discussion and Conclusions}
\label{sec:sec4}
We constructed broadband SEDs for three high-redshift BL Lac objects, J0022,
J0630, and J0811, using nearly contemporaneous observations in the
optical to X-ray band. Studying the LAT data, we conclude that the variability
on day to year timescales is fairly low for these three systems. This allows us to use the
6-year (mission averaged) LAT spectrum in forming our SED.
We fit the SEDs with a synchrotron/Compton model to infer physical properties of the sources.

Interestingly, Figure~\ref{fig:fig3} shows that there is a
trend for high-flux optical states to be spectrally harder.  Similar trends have
been seen in other blazars \citep[e.g.,][]{zlz+12}.
Our contemporaneous data (and SED modeling) are for the low, relatively quiescent state. We
lack the broad-band high state coverage to study the physical properties imposing this variation
via separate SED fits.
Still, if the variation (increase in $L^{\rm sy}_{\rm pk}$ and $\nu^{\rm sy}_{\rm pk}$)
were produced by an increase in the external photon field, one expects $\nu^{\rm sy}_{\rm pk}$
to decrease as the jet particles should cool more efficiently. This is not observed and so
we infer that the variation is likely produced in the injection particle spectrum or in the
jet blob flow (e.g., increase in $\delta_{\rm D}$) and $B$ field. This suggests correlated optical
GeV variability, which may be too weak for the LAT to detect.

        The basic B97 modeling constrains the emission parameters well under
the assumptions of pure SSC emission and radiative cooling of the injected
electrons (Figure~\ref{fig:fig3}). The SED fits assuming only the assigned statistical
errors is adequate (probabilities $pr = 10^{-2}$--$10^{-3}$)
However there are almost certainly additional
systematic errors including extinction uncertainty and inter-instrument calibrations.
For example, increasing the measurement uncertainties by 5\% (all the SED data points)
makes the fit acceptable, with $pr\sim$10\%.

        The SED parameters are, however, somewhat unusual, giving
particularly soft injection spectra, with $p_{\rm 1}$ well above that expected for relativistic
shock acceleration, $p_{\rm 1}\sim 2-2.5$.  For J0022 and J0811,
higher $p_{\rm 1}$ are required because of
the flat SED ($\alpha=0$ in $\nu F_{\rm \nu}\propto \nu^{\alpha}$)
in the optical band, which requires $p_{\rm 1}\sim3$. If we identify this with the cooled
spectrum, allowing harder injection, then we cannot accommodate the steeper X-ray spectrum
since radiative cooling produces only a $\Delta \alpha=0.5$ break
(if the electrons were in the  Klein-Nishina regime the break would be even weaker). Similarly,
matching the J0630 optical spectrum ($\alpha \sim 0.2$) and X-ray spectrum ($\alpha \sim -1$)
is not possible if we let the electrons cool with the break between the optical and
the X-ray bands (Figure~\ref{fig:fig4}). Thus we are forced to very steep injection spectra
if the X-rays are produced by the same population as the optical emission.
This conclusion is supported by fitting with more complex heuristic electron spectra (FDB08 model).
With such models we can avoid the very high magnetic field strength
required for J0630 to implement the rapid X-ray cooling and use lower
10\,mG fields.

        The minimum electron energies for the sources are rather high. While these values
are not unusual when compared to those in other works \citep[e.g.,][]{tgg+10},
it is not clear what environments/conditions are required in the acceleration site
to achieve such high minimum electron energies and
further investigations are needed to tell whether or not
such values are realistic. Note that we do not
use the equipartition magnetic-field strength in our modeling,
and the particle energy is much larger than the
magnetic energy in our models. In particular, the inferred magnetic field strength for J0811
is very low compared to those for previously studied BL~Lacs
\citep[see][for example]{fdb08, tgg+10, zlz+12},
although there are several objects  in the literature with lower inferred $B$
(and lower magnetic-to-particle-energy ratio). As we already noted (Section~\ref{sec:sec3_3}),
it may be possible to find another solution with lower $\gamma_{\rm min}$ and higher $B$.
Covering the SED more completely will help to infer the parameters
more precisely. Nevertheless, the SED at the high-energy end is primarily
determined by the X-ray spectrum in our model, and thus our conclusion
on the EBL would not change.

        By excluding the X-rays from the SED fit we can indeed accommodate lower injection $p_{\rm 1}$,
but the cost is that the X-ray must be an independent, steep spectrum component. Heuristic
modeling with inferred stationary e$^+$e$^-$ spectra confirm that a very steep population is
needed to model the X-ray component. Thus a simple, single-zone SSC model with typical
particle acceleration spectra is inadequate. The additional ingredient may be a separate, steep
cooled jet population for the X-ray emission. There is some indication for separate X-ray/optical
components seen in the different absorption columns inferred from the two bands for J0630.
However other effects (e.g. adiabatic expansion cooling) may also be relevant.

        We find that the $\gapp$100\,GeV LAT points for our highest redshift source J0630 are generally
significantly over-predicted by our SED models and take this to be strong evidence of the
effect of EBL absorption. Standard EBL models do a good job of producing the observed spectral
cutoff, but high UV models are not satisfactory
\citep[see also][]{fermiEBL, HESSEBL}.
This conclusion is fairly robust, and
EBL absorption is still required if we allow the observed X-ray emission to be a separate jet
component. Introduction of EC components from faint (unobserved) disk emission
affects the shape of the LAT spectrum. In general the harder EC spectrum does not match the LAT data
and it is difficult to arrange components to mimic the high-energy cutoff; EBL
absorption is still preferred unless the synchrotron cutoff is extraordinarily
sharp. We can approximate this with
an abrupt cut-off in the electron energy distribution (Figure~\ref{fig:fig5}c), but
such a sharp feature is unlikely to be realized in physical
acceleration models.
Note that the effects of EBL absorption are not clearly visible
in the low redshift sources as expected in EBL models; optical depth at 50\,GeV for $z=0.7$
is only 0.08 estimated with the Dom{\'{\i}}nguez model in Table~\ref{ta:ta4}.

\begin{figure*}
\centering
\begin{tabular}{cc}
\hspace{-0.0 mm}
\includegraphics[width=3.2 in]{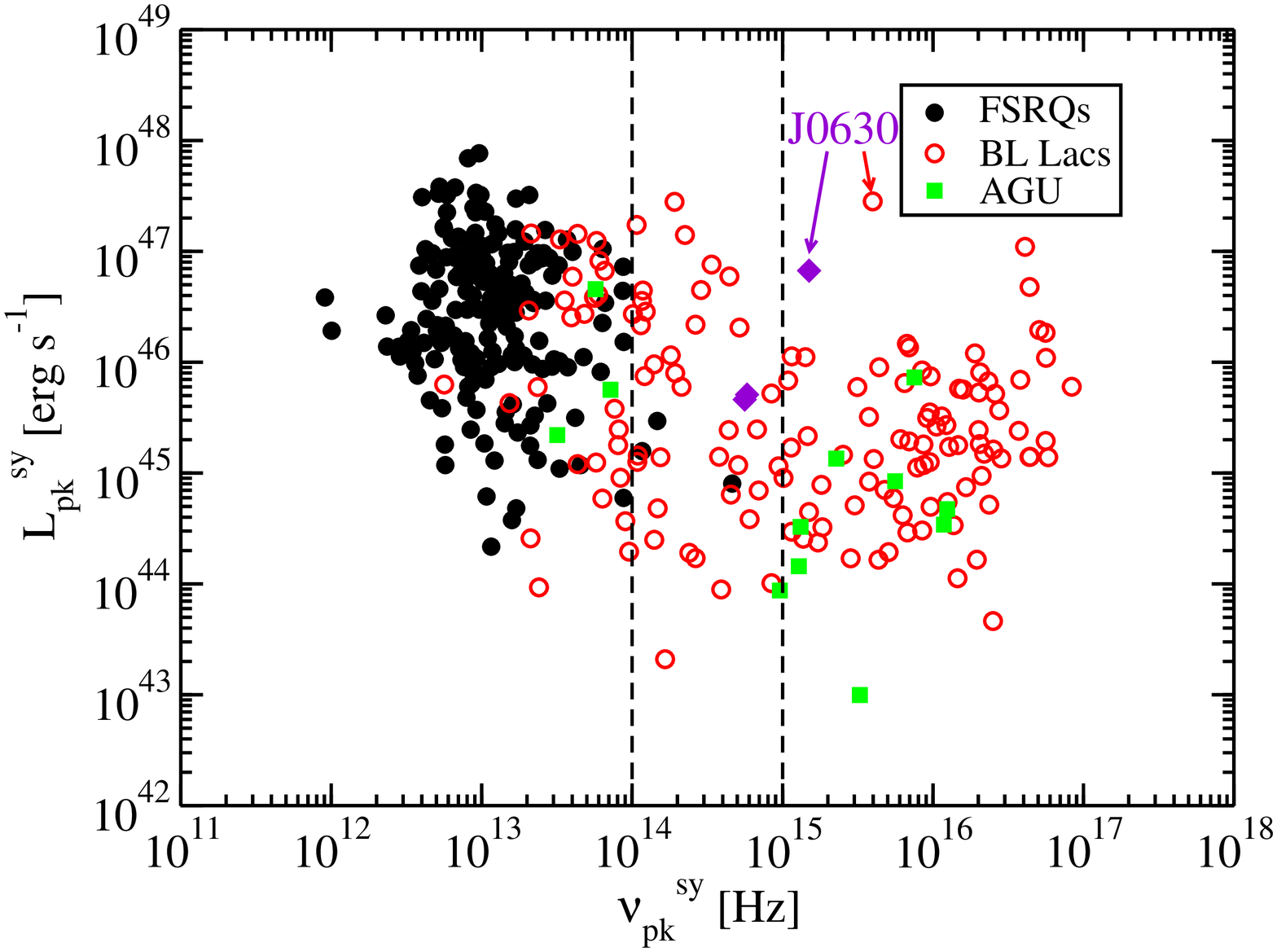} &
\hspace{4 mm}
\includegraphics[width=3.1 in]{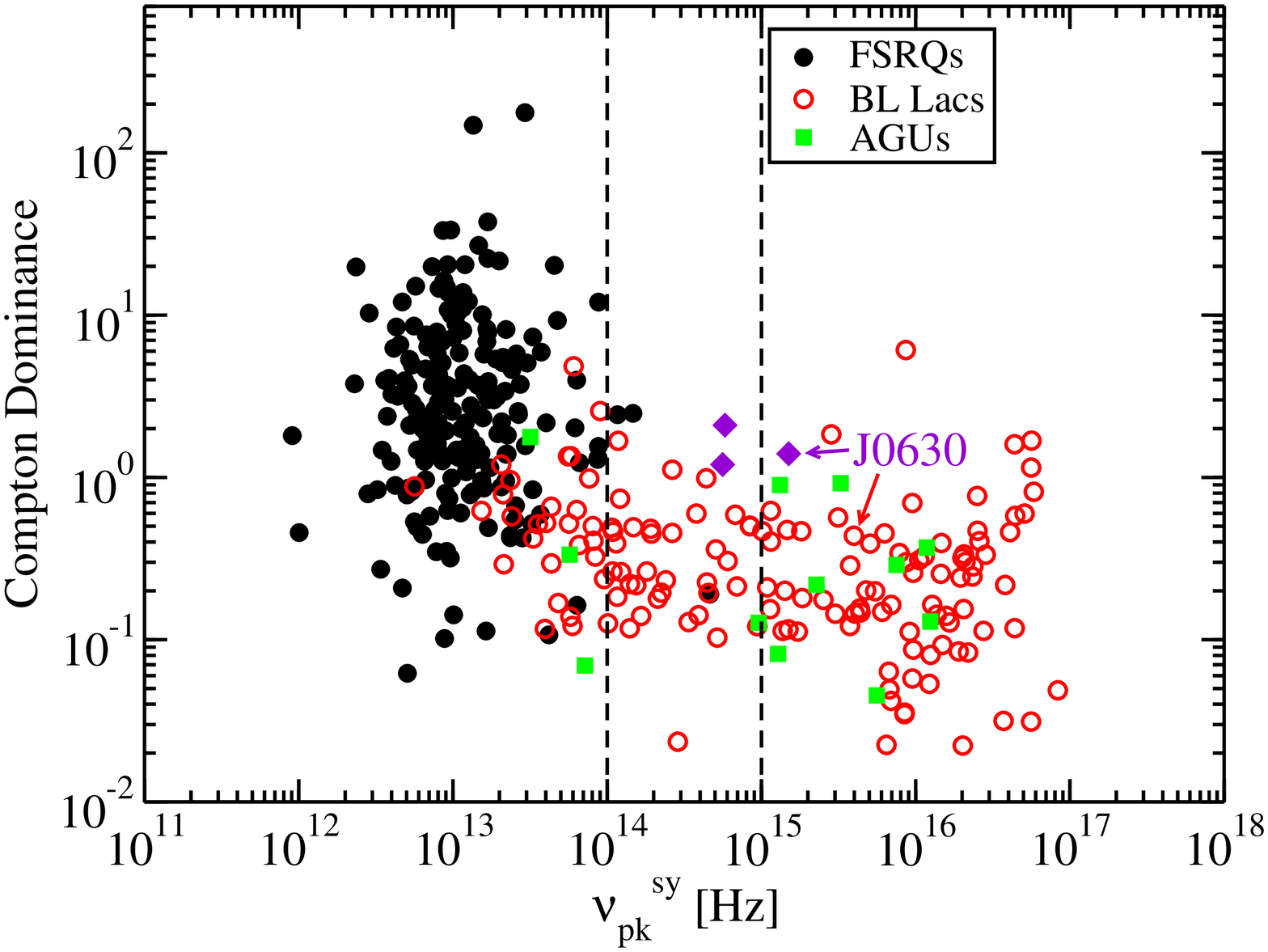} \\
\end{tabular}
\figcaption{{\it Left}: Synchrotron peak luminosity vs. synchrotron peak frequency.
{\it Right}: Compton dominance vs. synchrotron peak frequency.
We use black filled circles for FSRQs, red empty circles for BL Lacs, and green
squares for sources which are not clearly classified.
The three BL Lacs we study are shown as purple diamonds. Note that the red circle
for J0630 shows the position of the source reported in a previous study \citep[][]{f13}.
\label{fig:fig6}
}
\vspace{0mm}
\end{figure*}

We conclude with a few comments about the place of our sources in the BL Lac population.
Our objects are luminous with high $\nu_{\rm pk}^{\rm sy}$ so it is natural to consider
their relation to the `blazar sequence'.  In Figure~\ref{fig:fig6}, we plot $L^{\rm sy}_{\rm pk}$
and CD \citep[][]{f13} vs. $\nu_{\rm pk}^{\rm sy}$ (in the source rest frame) for
blazars from the 3LAC sample, including our three sources.
The general trend is commonly attributed to the effect of an increased external photon field
(e.g., from the BLR or disk) for blazars with lower $\nu^{\rm sy}_{\rm pk}$ and magnetic field strength
\citep[e.g.,][]{gcf+98, f13}.
Our three sources are HSPs/ISPs,
but are relatively close to the ISP border. They show higher $L_{\rm pk}^{\rm sy}$
and higher CD than the general population, but only J0630 is a true outlier, in the
$L_{\rm pk}^{\rm sy}$ plot. In fact with the quiescent state SED assembled here, it is
somewhat less extreme than in previous studies. Still, as one of the four high-redshift BL Lacs
called out by \citet[][]{pgr12} it does present some challenges to the simple blazar sequence.
A more complete study of the high-redshift LAT BL Lacs is needed to see if such sources are
a robust population and thus conflict with the blazar sequence correlation. If so, sources
such as J0630 may be FSRQs viewed very close to the jet axis ($\theta_{\rm v}<0.81$\,deg;
Table~\ref{ta:ta3}) so that the disk/BLR emission is overwhelmed by the beamed jet emission.
A detailed study along the lines of the blue FSRQ model \citep[][]{gtf+12}
using our high-quality contemporaneous SEDs
would be quite interesting.

        Since $L_{\rm pk}^{\rm sy}$ is redshift-dependent, it is more subject to
selection effects in a survey study. Thus it is argued \citep[e.g.,][]{f13} that CD is a more
robust classifier of the blazar status, being redshift independent (although still
sensitive to viewing angle effects, if EC components contribute). In
Figure~\ref{fig:fig6} right \citep[see][for more details]{f13}, we see that our three sources lie
near the upper edge of the HSP population. These are highly Compton-dominated sources but
not really distinct from the rest of the HSP population. Since our three sources, and the other
high-peak/high-power BL Lacs, still follow a general correlation in this plot, it suggests that
the blazar sequence scenario may still be robust to inclusion of high-power, high-redshift BL Lacs.

Nonetheless, the Doppler factors ($\delta_{\rm D}$) of these three sources are fairly large.
Following the cosmic evolution, \citet{arg+14} inferred the distribution of
the Lorentz factor ($\Gamma$) and the viewing angle ($\theta_{\rm v}$) for the LAT blazar
population. We note that the distribution for $\theta_{\rm v}$
derived by \citet{arg+14} (their Figure~9) is broad and the values we inferred with the
models (Tables~\ref{ta:ta3}) are not exceptional. However, the
best-fit Lorentz factors are very high considering the power-law distribution
with the slope $k=-2.03\pm0.70$ for BL Lacs \citep[][]{arg+14}. In order for
the chance probability of having $\Gamma>35.3$ (for J0630) to be greater than 1\%,
$k$ should be greater than $-2.49$. So perhaps our sources do represent a high velocity,
tightly beamed wing of the BL Lac population and their unusual properties are due to beaming
effects.

        Whether or not BL Lacs at $z>1$ contradict our present picture of the source evolution,
our SED measurements, particularly that for J0630, show that these sources can be a powerful
probe of the EBL and its evolution. We anticipate more striking EBL constraints, pushing to
the peak of cosmic star formation via further study of high-redshift {\it Fermi}-detected BL Lacs.

\bigskip

This work was supported under NASA Contract No. NNG08FD60C,
and  made use of data from the {\it NuSTAR} mission,
a project led by  the California Institute of Technology, managed by the Jet Propulsion  Laboratory,
and funded by the National Aeronautics and Space  Administration. We thank the {\it NuSTAR} Operations,
Software and  Calibration teams for support with the execution and analysis of  these observations.
This research has made use of the {\it NuSTAR}  Data Analysis Software (NuSTARDAS) jointly developed by
the ASI  Science Data Center (ASDC, Italy) and the California Institute of  Technology (USA).

The \textit{Fermi} LAT Collaboration acknowledges generous ongoing support
from a number of agencies and institutes that have supported both the
development and the operation of the LAT as well as scientific data analysis.
These include the National Aeronautics and Space Administration and the
Department of Energy in the United States, the Commissariat \`a l'Energie Atomique
and the Centre National de la Recherche Scientifique / Institut National de Physique
Nucl\'eaire et de Physique des Particules in France, the Agenzia Spaziale Italiana
and the Istituto Nazionale di Fisica Nucleare in Italy, the Ministry of Education,
Culture, Sports, Science and Technology (MEXT), High Energy Accelerator Research
Organization (KEK) and Japan Aerospace Exploration Agency (JAXA) in Japan, and
the K.~A.~Wallenberg Foundation, the Swedish Research Council and the
Swedish National Space Board in Sweden.

Additional support for science analysis during the operations phase is gratefully
acknowledged from the Istituto Nazionale di Astrofisica in Italy and
the Centre National d'\'Etudes Spatiales in France.

H.A. acknowledges supports provided by the NASA sponsored {\it Fermi}
Contract NAS5-00147 and by
Kavli Institute for Particle Astrophysics and Cosmology (KIPAC).
Part of the funding for GROND (both hardware as well as personnel)
was generously granted from the Leibniz-Prize to Prof. G. Hasinger
(DFG grant HA 1850/28-1).

\bibliographystyle{apj}
\bibliography{BLLacs}

\end{document}